\documentclass[conference]{IEEEtran}
\IEEEoverridecommandlockouts
\usepackage{cite}
\usepackage{amsmath,amssymb,amsfonts}
\usepackage{algorithmic}
\usepackage{algorithm}
\usepackage{graphicx}
\usepackage{bm}
\usepackage{textcomp}
\usepackage{hyperref}
\usepackage{xcolor}
\usepackage{subfigure}
\def\BibTeX{{\rm B\kern-.05em{\sc i\kern-.025em b}\kern-.08em
    T\kern-.1667em\lower.7ex\hbox{E}\kern-.125emX}}

\begin{document}
\title{Supporting More Active Users for Massive Access via Data-assisted Activity Detection\\
\thanks{This work was supported by the General Research Fund (Project No. 15207220) from the Hong Kong Research Grants Council.}
}

\author{\IEEEauthorblockN{Xinyu Bian, Yuyi Mao, and Jun Zhang}
\IEEEauthorblockA{{Department of Electronic and Information Engineering} \\
{The Hong Kong Polytechnic University}, Hong Kong \\
Emails: xinyu.bian@connect.polyu.hk, yuyi-eie.mao@polyu.edu.hk, jun-eie.zhang@polyu.edu.hk}
}

\maketitle
\begin{abstract}
Massive machine-type communication (mMTC) has been regarded as one of the most important use scenarios in the fifth generation (5G) and beyond wireless networks, which demands scalable access for a large number of devices. While grant-free random access has emerged as a promising mechanism for massive access, its potential has not been fully unleashed. Particularly, the two key tasks in massive access systems, namely, user activity detection and data detection, were handled separately in most existing studies, which ignored the common sparsity pattern in the received pilot and data signal. Moreover, error detection and correction in the payload data provide additional mechanisms for performance improvement. In this paper, we propose a data-assisted activity detection framework, which aims at supporting more active users by reducing the activity detection error, consisting of false alarm and missed detection errors. Specifically, after an initial activity detection step based on the pilot symbols, the false alarm users are filtered by applying \emph{energy detection} for the data symbols; once data symbols of some active users have been successfully decoded, their effect in activity detection will be resolved via \emph{successive pilot interference cancellation}, which reduces the missed detection error. Simulation results show that the proposed algorithm effectively increases the activity detection accuracy, and it is able to support $\sim 20\%$ more active users compared to a conventional method in some sample scenarios.
\end{abstract}
  
\begin{IEEEkeywords}
Internet-of-Things (IoT), massive connectivity, grant-free massive access, data-assisted user activity detection, approximate message passing (AMP).
\end{IEEEkeywords}

\section{Introduction}
The proliferation of the Internet of Things (IoT), such as connected health, smart home, and intelligent manufacturing, is prompting a rapid revolution of wireless communications. In order to support a massive number of connected devices, massive machine-type communications (mMTC) has become one of the three generic services offered by the fifth generation (5G) wireless networks \cite{b1}. A unique feature of mMTC is that, while a huge amount of devices are connected, only a proportion of them sporadically become active, normally with a small amount of data to transmit \cite{b2}. 

Nevertheless, uplink access in legacy wireless networks is generally controlled by grant-based access schemes, where each user first transmits a scheduling request to the base station (BS) and cannot start its data transmission until a grant is received. Although the grant-based access schemes reserve dedicated resources for each user that avoids potential collisions, long latency and significant signalling overhead will be incurred with a large number of devices \cite{b3,b4}.

Grant-free random access, where users can transmit data without waiting for approval from the BS \cite{b5}, provides a promising solution for mMTC. In its protocols, the BS needs to detect the set of active users and estimate their channel conditions based on the received pilot signal, before performing data reception operations. Due to the vast amount of devices, users can only be assigned with non-orthogonal pilots, which makes it highly challenging for accurate active user identification and channel estimation at the BS. As a result, accommodating the maximum number of active devices with minimum degradation of communication performance is widely acknowledged as one of the most fundamental design considerations for grant-free massive access \cite{b3,b6,b7}.

Because of the sporadic traffic pattern of the connected devices, detecting the set of active users turns out to be a compressive sensing problem, for which, many efficient algorithms were developed \cite{b8}. In \cite{b9}, a joint user activity detection and data detection algorithm was proposed for grant-free non-orthogonal multiple access (NOMA) by exploiting the temporal correlations of user activities. A similar problem was later revisited using approximate message passing (AMP) and expectation maximization (EM) in \cite{b10}. However, these works assume full channel state information (CSI) available at the BSs, which is practicallly infeasible since most of the users are inactive without transmitting their pilots to the BS. Therefore, joint activity detection and channel estimation has attracted significant attentions most recently \cite{b11,b12}. In \cite{b11}, a joint design of activity detection and channel estimation was proposed based on AMP for massive multi-input multi-output (MIMO) systems, and it was shown that the activity detection error can be arbitrarily small in the asymptotic regime. In addition, a user activity detection and channel estimation approach was developed in \cite{b12} by leveraging the joint sparsity from both the spatial and frequency domains. This approach obviates the need of knowing the number of devices.

However, prior works on grant-free massive access mostly follow a separate design approach, i.e., the activity pattern and CSI are estimated without incorporating any information encoded in the received data symbols. In this way, it only utilizes the sparse activity pattern from the received pilot signal, which limits the activity detection accuracy and the data transmission reliability. An important but easily neglected observation in grant-free random access is that the same user activity pattern replicates in the received data symbols, which can be exploited to improve the activity detection accuracy for accommodating more connected devices. This inspires the design of a data-assisted activity detection framework in this paper, where the false alarm and missed detection error can both be suppressed. It is worthwhile to note that this idea was initially proposed for a single-antenna NOMA-based massive access system \cite{b13}, which, however, cannot be easily extended for multi-antenna receptions.

In this paper, we endeavor to reduce the activity detection error by leveraging valuable information obtained in data symbols. The proposed data-aided activity detection framework contains three basic modules, namely, an initial estimator, a false alarm corrector and a missed detection corrector. On one hand, to minimize the false alarm error, energy detection is applied in the false alarm corrector to filter inactive users that are incorrectly determined as active. On the other hand, inspired by the successive interference cancellation (SIC) detection, the missed detection corrector progressively increases the sparsity level of the received pilot signal to reduce the probability of missed detection. Simulation results shows that the proposed framework is able to achieve noticeable improvements in terms of both user activity detection accuracy and data detection error. Moreover, about 20\% more active users can be supported by the proposed framework in sample scenarios, compared to that achieved by the separate design.

The rest of this paper is organized as follows. We introduce the system model and two basic tasks of grant-free access in Section \uppercase\expandafter{\romannumeral2}. A data-assisted user activity detection framework is developed in Section \uppercase\expandafter{\romannumeral3}. Simulation results are presented in Section \uppercase\expandafter{\romannumeral4}, and we conclude this paper in Section \uppercase\expandafter{\romannumeral5}.

\textbf{Notations:} We use lower-case letters, bold-face lower-case letters, bold-face upper-case letters, and math calligraphy letters to denote scalars, vectors, matrices, and sets, respectively. Besides, the conjugate transpose of a matrix $\mathbf{M}$ is denoted as $\mathbf{M}^{H}$ and the complex Gaussian distribution with mean $\bm{\mu}$ and covariance matrix $\bm{\Sigma}$ is denoted by $\mathcal{C} \mathcal{N}(\bm{\mu}, \bm{\Sigma})$. In addition, the indicator function and the Kronecker product are denoted as $\mathbf{1}(\cdot)$ and $\otimes$, respectively. We use $\text{vec}(\cdot)$ to denote the vectorization operator and let $\text{vec}^{-1}(\cdot)$ denote its inverse.

\section{System Model}
\subsection{Signal Model}
\begin{figure}[htpb]
\centering
\includegraphics[height=4.2cm,width=7.71826cm]{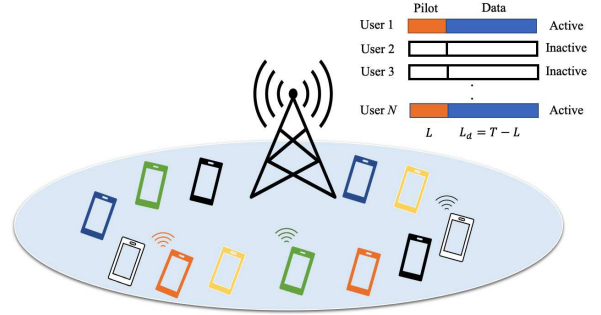}
\caption{System model and the adopted grant-free random access scheme.}
\label{fig1}
\end{figure}

We consider an uplink cellular system as shown in Fig. \ref{fig1}, where a large number of mobile users are simultaneously served by a BS. The scenarios where the mobile users have sporadic uplink data traffic (e.g., the IoT and mMTC) are of particular interests, where only a small fraction of the users have data to transmit and become active at each time instant. The active probabilities of different users are assumed to be identical, and they are denoted as $p$. We denote the set of mobile users as $\mathcal{N}\triangleq \{1,\cdots,N\}$, and use the activity indicator $u_{n}\in\{0,1\}$ to represent whether a user is active for transmission, i.e., $u_{n}=1$ indicates the user is active and $u_{n} = 0$ if it is inactive. The set of active users is represented by $\Xi \triangleq \left\{j \in \mathcal{N} | u_{j}=1 \right\}$ with its cardinality denoted as $K$ ($K\leq N$). For simplicity, the BS is assumed to have $M$ receive antennas while each user transmits with a single antenna.

We adopt the quasi-static block fading channel model, where the channel condition remains unchanged within a transmission block spanning $T$ symbol intervals, and changes independently across different coherence blocks. The uplink channel vector from user $n$ to the BS, denoted as $\mathbf{h}_{n}$, is modeled as $\mathbf{f}_{n}=\sqrt{\beta_{n}} \boldsymbol{\alpha}_{n}, \forall n$, where $\boldsymbol{\alpha}_{n}$ and $\beta_{n}$ stand for the small-scale and large-scale fading coefficients, respectively. Besides, the users are assumed to be static and thus $\beta_n$ is known at the BS.

A grant-free random access scheme as shown in Fig. \ref{fig1}, is adopted for uplink transmissions, where a transmission block is divided into two phases: The first phase contains $L$ symbols that are reserved for pilot transmission and the remaining $L_{d}\triangleq T-L$ symbols are used for payload data delivery in the second phase. We consider the massive random access scenarios, i.e., $L<N$, in which, assigning orthogonal pilot sequences to all the users is infeasible. To overcome this issue, each user is instead assigned with a unique pilot sequence $\sqrt{L}\mathbf{a}_{n}$ with $\mathbf{a}_{n}\triangleq \left[a_{n, 1}, \cdots, a_{n, L}\right]^{T}$ and $a_{n, l} \sim \mathcal{C} \mathcal{N}\left(0, \frac{1}{L}\right)$ \cite{b7}. It can be verified that $\{\mathbf{a}_{n}\}_{n=1}^{N}$ achieves asymptotic orthogonality when $L$ is sufficiently large. By defining $\mathbf{A}_{p}$ as $\left[\mathbf{a}_{1}, \cdots, \mathbf{a}_{N}\right]$, the received pilot signal $\mathbf{Y}_{p} \in \mathbb{C}^{L\times M}$ at the BS in the first phase can be expressed as follows: 
\hspace{1pt}
\begin{align}
\mathbf{Y}_{p}=\sqrt{L\rho}\mathbf{A}_{p}\mathbf{H}+\mathbf{N}_{p},
\end{align}
\noindent where $\rho$ is the user transmit power, $\mathbf{H} \triangleq \left[\mathbf{h}_{1},...,\mathbf{h}_{N}\right]^{T}$ denotes the effective channel matrix with $\mathbf{h}_{n}\triangleq u_{n}\mathbf{f}_{n}$, and $\mathbf{N}_{p}=\left[\mathbf{n}_{p,1},...,\mathbf{n}_{p,L}\right]^{T}$ is the Gaussian noise with zero mean and variance $\sigma^{2}$ for each element.

In the data transmission phase, each active user transmits $s$ ($s < L_d$) coded symbols, which is denoted as $\mathbf{s}_{n} \in \mathcal{X}^{s \times 1}$. Here, $\mathcal{X}$ is the set of constellation points with the normalized average power. For the set of inactive users, $\mathbf{s}_{n}$ is set to be a zero vector for notation consistency. Since the number of active users in the system may far exceed the number of receive antennas at the BS, in order to avoid the system from being overloaded \cite{b14}, we multiply the coded symbols by a precoding matrix for each user \cite{b15} as follows \vspace{1pt}
\begin{align}
\mathbf{c}_{n} &=\mathbf{P}_{n}\mathbf{s}_{n},
\end{align}

\noindent where $\mathbf{c}_n$ is the precoded symbols and $\mathbf{P}_{n} \in \mathbb{C}^{L_{d} \times s}$ is the precoding matrix with full column-rank. Thus, the received data signal at the BS, denoted as $\mathbf{Y}_{d}\in \mathbb{C}^{M \times L_{d}}$, can be expressed as follows: 
\hspace{1pt}
\begin{align}
\mathbf{Y}_{d}&=\sqrt{\rho} \sum_{n=1}^{N} \mathbf{h}_{n} \mathbf{c}_{n}^{T}+\mathbf{N}_{d} =\sqrt{\rho} \sum_{j \in \Xi} \mathbf{h}_{j} \mathbf{s}_{j}^{T}\mathbf{P}_{j}^{T}+\mathbf{N}_{d},
\end{align}

\noindent where $\mathbf{N}_{d}=\left[\mathbf{n}_{d,1},...,\mathbf{n}_{d,L_{d}}\right]$ is the Gaussian noise with the same distribution as $\mathbf{N}_{p}$. We denote $\mathbf{y}_{d} = \text{vec}\left(\mathbf{Y}_{d}\right)$, and let $\mathbf{B}_{n}\triangleq \mathbf{P}_{n} \otimes \mathbf{h}_{n}$. As a result, the received data signal in (3) can be rewritten as the following expression: 
\hspace{1pt}
\begin{align}
\mathbf{y}_{d}=\sqrt{\rho} \mathbf{B}_{a} \mathbf{x}_{a}+\mathbf{N}_{d},
\end{align}

\noindent where $\mathbf{B}_{a} \triangleq \left[ \{\mathbf{B}_{j}\}_{ j \in \Xi} \right] $ and $\mathbf{x}_{a} \triangleq \left[\{\mathbf{s}_{j}^{T}\}_{j \in \Xi} \right]$.

\subsection{User Activity and Data Detection}
\emph{User activity detection} and \emph{data detection} are the two most critical tasks in grant-free massive access. Prior studies on massive connectivity typically adopted a two-stage separate design as shown in Fig. \ref{fig2a} \cite{b10,b16,b17}. Specifically, in the first stage, activity detection and channel estimation are performed based on the received pilot signal, which can be accomplished by exploiting the sparsity of the effective channel matrix using compressive sensing techniques \cite{b3}. The estimated user activity pattern and CSI are then used for data detection in the second stage.

With limited resources available for pilot transmissions, it is challenging to obtain accurate knowledge of the user activity pattern at the BS. In fact, missed detection, i.e., an active user is not detected at all, and false alarm, i.e., an inactive user is determined as active, are two major sources that contribute to the user activity detection error. On one hand, data of the miss-detected users is not decoded, leading to a one-hundred percent data error for these users; On the other hand, false alarm shall degrade the data detection accuracy, since the data detector also attempts to decode data for the false alarm users, which is equivalent to introducing interference to the active users. Therefore, improving the activity detection accuracy is of the utmost importance to the communication performance in massive access systems.

A key observation of the grant-free access scheme is that, both the transmitted pilots and data symbols are distorted by the same wireless fading channel. In other words, the received pilot and data signals share the same sparsity pattern, which could be exploited to improve the activity detection accuracy. Nevertheless, this aspect was largely overlooked by existing studies, which motivates our investigation on data-assisted activity detection approaches. In the next section, we will customize dedicated methods to handle the two kinds of errors, in order to reduce the overall user activity detection error for reliable communications.

\section{The Proposed Framework}
\begin{figure} 
  \centering 
  \subfigure[Separate design of user activity and data detection.]{ 
    \label{fig2a} 
    \includegraphics[height=2cm,width=7.93717277cm]{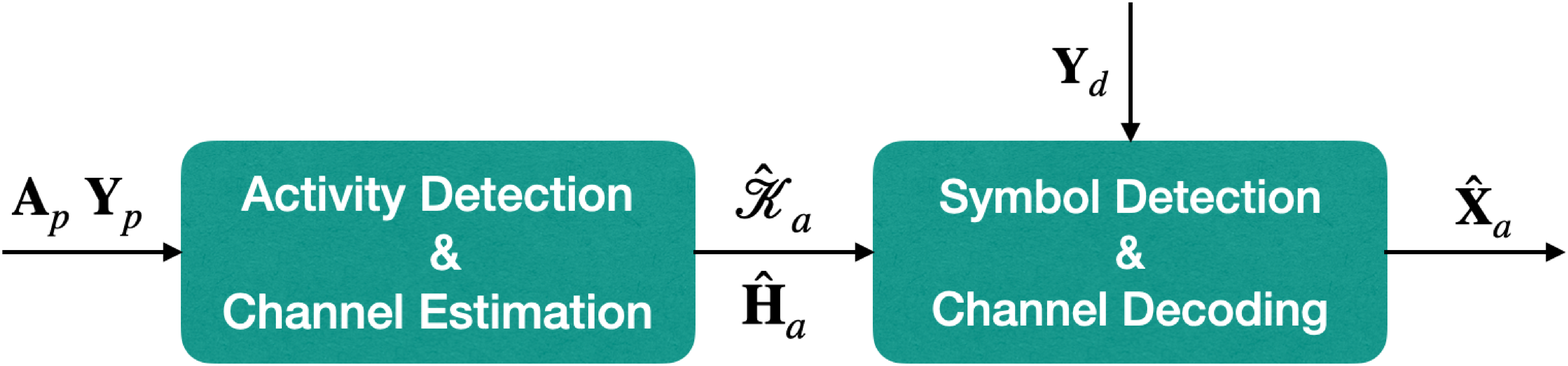}} 
  \hspace{1in} 
  \subfigure[The data-assisted activity detection framework.]{ 
    \label{fig2b} 
    \includegraphics[height=6cm,width=9.02471169cm]{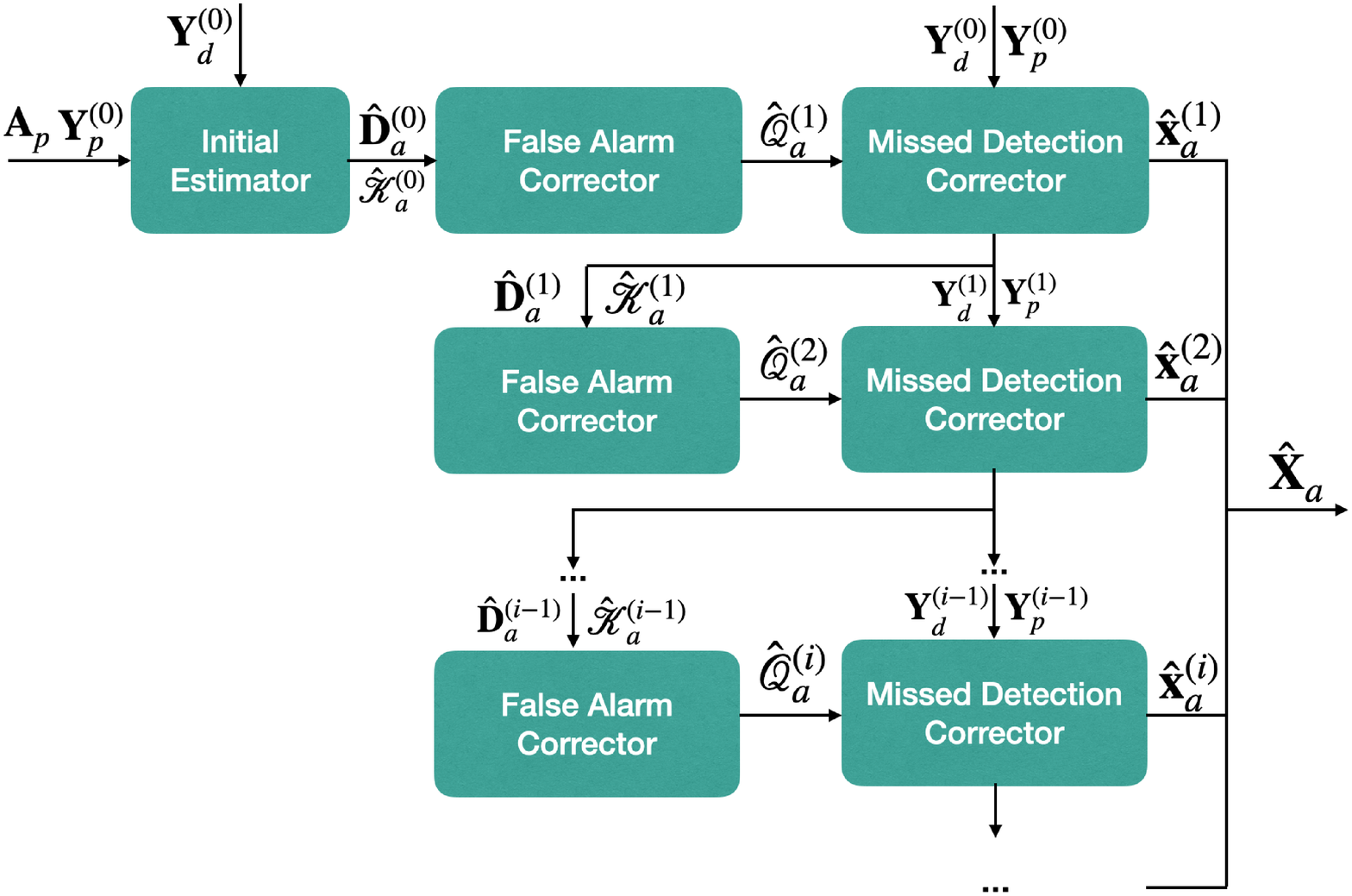}} 
  \caption{The separate design and data-assisted design for massive access.} 
  \label{fig2} 
\end{figure}

In this section, we propose a data-assisted activity detection framework to improve the activity detection accuracy. A flow chart of the proposed framework is shown in Fig. \ref{fig2b}, which contains an initial estimator, a false alarm corrector, and a missed detection corrector. For each transmission block, the initial estimator performs preliminary estimation on the CSI and the user activity pattern for subsequent data detection. This is essentially the separate design as shown in Fig. \ref{fig2a}. Based on the initial data symbol estimates, the false alarm corrector performs energy detection to filter part of the false alarm users. This step is inspired by the intuition that the average magnitudes of the detected data symbols of the false alarm users shall be much smaller than those of the active users. Then, with the updated user activity pattern, the channel matrices and data symbols are re-estimated for processing in the missed detection corrector. In particular, the design of the missed detection corrector leverages the SIC techniques to further refine the activity detection result. In contrast to conventional SIC algorithms that remove interference from the received data signal, interference in the received pilot signal is eliminated by identifying a subset of users whose payload data can be successfully decoded in each iteration.

We will elaborate different modules of the proposed framework in the following subsections. For better expositions, we use the superscript ``$^{(i)}$'' to denote the iteration number, and refer the operations of the initial estimator as the $0$-th iteration. Besides, the intermediate variables $N^{(0)}$, $K^{(0)}$, $\mathbf{Y}_{p}^{(0)}$, $\mathbf{y}_{d}^{(0)}$ and $\mathcal{N}^{(0)}$ are initialized as $N$, $K$, $\mathbf{Y}_{p}$, $\mathbf{y}_{d}$ and $\mathcal{N}$, respectively.

\subsection{The Initial Estimator}
The initial estimator applies the AMP-based algorithm proposed in \cite{b10} to jointly estimate the CSI and user activity pattern, based on which, data symbol detection is performed. In particular, with $\mathbf{Y}_p^{(i)}$ as the input\footnote{In this subsection, we retain the superscript ``$^{(i)}$" as the key steps in the initial estimator that are reused in the missed detection corrector as will be discussed in Section III-D, in which the iteration number is greater than zero.}, the AMP-based algorithm obtains the channel estimates for the users in $\mathcal{N}^{(i)}$, denoted as $\hat{\mathbf{H}}^{(i)}=\left[ \{\hat{\mathbf{h}}_{j}^{(i)}\}_{j \in \mathcal{N}^{(i)}} \right]$, and the user activity pattern is derived by thresholding, i.e., the set of active users is determined as $\hat{\mathcal{K}}_{a}^{(i)} \triangleq \left\{j \in \mathcal{N}^{(i)} | \phi(\hat{\mathbf{h}}_{j}^{(i)}) \geq {\theta}_{j}^{(i)} \right\}$, where $\phi(\cdot)$ is a known function and ${\theta}_{j}^{(i)}$ is the decision threshold for user $j$. 

We define $\hat{\mathbf{H}}_{a}^{(i)} \triangleq \left[\{ {\hat{\mathbf{h}}_{j}^{(i)}}\}_{ j \in \hat{\mathcal{K}}_{a}^{(i)}} \right]$ and $\hat{\mathbf{B}}_{a}^{(i)}$ in a way similar to $\mathbf{B}_{a}$ in (4). By using the MMSE equalizer, the estimated data symbols for the users in $\hat{\mathcal{K}}_{a}^{(i)}$, are obtained via the following expression: 

\begin{align}
\hat{\mathbf{D}}_{a}^{(i)}=\text{vec}^{-1} \left[ \left(\hat{\mathbf{B}}_{a}^{(i)H} \hat{\mathbf{B}}_{a}^{(i)}+\frac{\sigma^{2}}{\rho} \mathbf{I}\right)^{-1} \hat{\mathbf{B}}_{a}^{(i)H}\mathbf{y}_{d}^{(i)} \right],
\end{align}

\noindent where $\hat{\mathbf{D}}_{a}^{(i)}\triangleq \left[ \{ \hat{\mathbf{d}}_{a,j}^{(i)}\}_{j \in \hat{\mathcal{K}}_{a}^{(i)}} \right]$ with $\hat{\mathbf{d}}_{a,j}^{(i)} \triangleq \left[\{\hat{d}_{a,(j,m)}^{(i)}\}_{m=1}^{s} \right]$ and $\hat{d}_{a,(j,m)}^{(i)}$ is the $m$-th estimated data symbol of user $j$.

\subsection{The False Alarm Corrector}
In the $i$-th iteration, the false alarm corrector filters the inactive users from $\hat{\mathcal{K}}_{a}^{(i-1)}$, which is obtained from the initial estimator if $i=1$ and the missed detection corrector in the previous iteration if $i\geq 2$. We borrow the idea of energy detection for spectrum sensing in cognitive radio networks \cite{b18} to design the false alarm corrector. This is because if the estimated data symbols of a user have small average magnitudes, this user is likely to be a false alarm user. 

Specifically, in the false alarm corrector, a user that was detected as active in the previous iteration is determined as a false alarm user if the following criteria is satisfied: 
\hspace{1pt}
\begin{align}
\sum_{m=1}^{s} \mathbf{1} \left(|\hat{d}_{a,(j,m)}^{(i-1)}| \in (0, \theta_{F_{1}}) \cup (\theta_{F_{2}}, +\infty) \right)\geq\theta_{F_{3}}, \forall j \in \hat{\mathcal{K}}_{a}^{(i-1)}.
\end{align}

\noindent In (6), $\theta_{F_{1}}$, $\theta_{F_{2}}$ and $\theta_{F_{3}}$ are empirical threshold values, where $\theta_{F_{1}}$ is to ensure the average estimated data symbol energy of an active user is sufficiently large, while $\theta_{F_{2}}$ is designed for reducing the sensitivity of the false alarm corrector to the channel estimation error. Therefore, the updated estimate of the user activity pattern is given by $\hat{\mathcal{Q}}_{a}^{(i)} \triangleq \hat{\mathcal{K}}_{a}^{(i-1)} \backslash$ $\big\{$$j \in \hat{\mathcal{K}}_{a}^{(i-1)} | \sum_{m=1}^{s} \mathbf{1} \left( |\hat{d}_{a,(j,m)}^{(i-1)}|\in (0, \theta_{F_{1}}) \cup (\theta_{F_{2}}, +\infty)\right)\geq\theta_{F_{3}}$\!$\big\}$.

\subsection{The Missed Detection Corrector}
\subsubsection{\textbf{Overview}}
While the false alarm corrector is able to reduce the chances of including inactive users in $\hat{\mathcal{K}}_a^{(i-1)}$, it cannot effectively handle the missed detection users. In this subsection, we design a missed detection corrector to minimize the number of active users that cannot be found in previous steps. Our design is motivated by the SIC techniques for multi-user detection \cite{b19}, where data from different users are detected sequentially, and interference in the received data signal is iteratively removed for decoding data of the remaining users. However, as our objective is to reduce the missed detection error, we propose to perform SIC for the received pilot signal instead in the missed detection corrector. By identifying some users that are determined as active with high confidence and remove their pilot data from the received pilot signal in each iteration, we shall be able to increase the sparsity level of the received pilot signal, which is beneficial for accurate user activity detection and channel estimation in next iterations.

\begin{figure}[htpb]
\centering
\includegraphics[height=2.6cm,width=9.10601848cm]{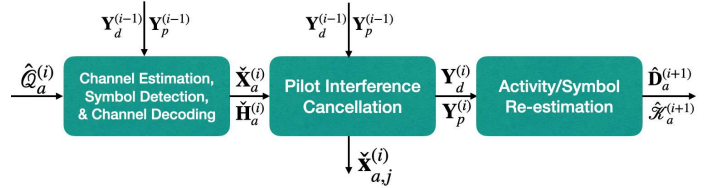}
\caption{The structure of the missed detection corrector.}
\label{fig3}
\end{figure}

In particular, the missed detection corrector performs three tasks as shown in Fig. \ref{fig3}, including \emph{i) Channel estimation, data symbol detection and channel decoding}; \emph{ii) Pilot interference cancellation}; and \emph{iii) User activity/symbol re-estimation}. These tasks will be elaborated in the sequel.

\subsubsection{\textbf{Channel estimation, symbol detection, and channel decoding}} 
With the updated estimate of the active user set $\hat{\mathcal{Q}}_{a}^{(i)}$ from the false alarm corrector, the missed detection corrector first re-estimates the channel vectors and performs data symbol detection accordingly, both of which apply the MMSE estimators. The estimated channel vectors of user $j$ is denoted as $\check{\mathbf{h}}_{j}^{(i)}$, and the detected symbol sequence on the constellation, i.e., which constellation points are transmitted in the symbol sequence, is denoted as $\check{\mathbf{s}}_{a,j}^{(i)}$ for user $j$. The detected symbol sequence is then passed to a channel decoder, which outputs the parity check result in addition to the decoded data. We denote the set of users that pass the parity check as $\hat{\mathcal{P}}_a^{(i)}$, whose channel-decoded data is denoted as $\check{\mathbf{x}}_{a,j}^{(i)}$, $j \in \hat{\mathcal{P}}_a^{(i)}$.

\subsubsection{\textbf{Pilot interference cancellation and activity/symbol re-estimation}}
After channel decoding, the missed detection corrector selects a number of users from $\hat{\mathcal{Q}}_a^{(i)}$ based on their parity check results. The pilots of the selected set of users are then subtracted from the received pilot signal. Our heuristics originate from a key theorem in compressive sensing (See Theorem 1.3 in \cite{b20}). This theorem implies that for an idealized user activity detection problem where the BS has a single receive antenna and without the receive noise, if $t$ ($t\leq K$) of the active users can be identified by an oracle, the remaining $K-t$ active users can also be accurately identified from the interference-cancelled pilot signal as long as the triplet $(N, K, L)$ satisfies the following inequality: 

\begin{align}
L \geq C(K-t)\ln\left(\frac{N-t}{K-t}\right),  t = 0, 1, 2, \cdots, K,
\end{align}

\noindent where $C>0$ is a constant. Since the right-hand side of (7) decreases with $t$, it means that if more active users can be accurately found by an oracle, the perfect active user recovery condition can be met more easily for the remaining users. In other words, increasing the sparsity level in the received pilot signal is useful to improve the activity detection performance. Unfortunately, this conclusion is drawn by imposing strict assumptions, which is rarely the case in practice.

To resolve this issue, the missed detection corrector selects $\min\{S_{a}, |\hat{\mathcal{P}}_a^{(i)}|\}$ users from $\hat{\mathcal{Q}}_a^{(i)}$ based on the channel decoding results, and the set of selected users for pilot interference cancellation in the $i$-th iteration, is denoted as $\hat{\mathcal{M}}_a^{(i)}$. Here, $S_{a}$ is a preset parameter in the proposed framework. In case that $|\hat{\mathcal{P}}_a^{(i)}| > S_{a}$, the $S_a$ users with the minimum Euclidean distances between $\check{\mathbf{x}}_{a,j}^{(i)}$ and $\check{\mathbf{s}}_{a,j}^{(i)}$ are selected. Thereafter, the pilot data of the selected set of users is subtracted from the received pilot signal $\mathbf{Y}_{p}^{(i-1)}$ for the next iteration:

\begin{align}
\mathbf{Y}_{p}^{(i)}=\mathbf{Y}_{p}^{(i-1)}-\sqrt{L\rho}\sum\nolimits_{j \in \hat{\mathcal{M}}_a^{(i)}} {\check{\mathbf{h}}_{j}^{(i)}} \mathbf{a}_{j}^{T},
\end{align}

\noindent and $\mathbf{Y}_{d}^{(i)}$ is updated accordingly as follows:

\begin{align}
\mathbf{Y}_{d}^{(i)}=\mathbf{Y}_{d}^{(i-1)}-\sqrt{\rho}\sum\nolimits_{j\in \hat{\mathcal{M}}_a^{(i)}} {\check{\mathbf{x}}_{a,j}^{(i)}} \check{\mathbf{h}}_{j}^{(i)}.
\end{align}

\noindent If $\hat{\mathcal{P}}_a^{(i)}=\emptyset$, the proposed algorithm will be terminated and the data decoding results of all users from $\hat{\mathcal{Q}}_a^{(i)}$ are deemed as incorrect. Otherwise, the AMP-based algorithm adopted by the initial estimator will be invoked again with $\mathbf{Y}_{p}^{(i)}$, $\mathbf{Y}_{d}^{(i)}$, $N^{(i)}=N^{(i-1)}-|\hat{\mathcal{M}}_a^{(i)}|$ and $K^{(i)}=K^{(i-1)}-|\hat{\mathcal{M}}_a^{(i)}|$ as input for re-estimating the user activity pattern and data symbols before calling the false alarm corrector in the next iteration. 

\section{Simulation Results}
\subsection{Simulation Settings and Baseline Schemes}
We simulate a single-cell uplink cellular network with $N=500$ users to corroborate the effectiveness of the proposed data-assisted user activity detection algorithm, where the users are located on a circle with a radius of 500 m to the BS. Each element of the precoding matrix $\mathbf{P}_{n}$ is sampled from the complex Gaussian distribution with zero mean and unit variance. In addition, we apply an idealized channel coding scheme, where perfect data recovery is assumed to be feasible if the symbol error in each block is below 20\%. It is worthy mentioning that the proposed algorithm is readily applicable for practical channel coding schemes, such as the low-density parity-check codes (LDPC) \cite{b21}. The simulation results are averaged over $10^7$ independent realizations. Other critical parameters used in simulations are summarized in TABLE \ref{table1}.

\begin{table}[ht]
\caption{Simulation parameters}
\centering
    \begin{tabular}{c|c|c|c}
    \hline
     Parameters & Values & Parameters & Values\\
    \hline
    $M$ & 32 & $L$ & 100\\
    $L_{d}$ & 70 & $s$ & 5\\
    $\beta_{n}$ & $-116.78$ dB & $\boldsymbol{\alpha}_{n}$ & $\mathcal{C} \mathcal{N}(\mathbf{0}, \mathbf{I}_{M})$\\
    $\theta_{F_{1}}$ & 0.2 & $\theta_{F_{2}}$ & 2\\
    $\theta_{F_{3}}$ & 3 & $S_a$ & 20\\
    Bandwidth & 1 MHz & Modulation & QPSK\\
    Transmit power & 23 dBm & Noise power density & -169 dBm/Hz\\
    \hline
    \end{tabular}
\label{table1}
\end{table}

We adopt three baseline schemes for comparisons:

\begin{itemize}
    \item \textbf{Separate design:} This scheme was proposed in \cite{b10}, where channel estimation and user activity detection are first performed using the AMP-based algorithm as stated in Section III-A. After that, data symbols are detected using an MMSE estimator.
\end{itemize}

\begin{itemize}
    \item \textbf{The proposed algorithm with false alarm correction only:} The main purpose of comparing this scheme is to reveal the impacts of the false alarm users on the system performance, including the activity detection error and data error. Specifically, we execute the proposed framework without invoking the missed detection corrector in this baseline scheme.
\end{itemize}

\begin{itemize}
    \item \textbf{Perfect knowledge of the user activity pattern:} This scheme assumes perfect knowledge of the user activity pattern and consequently, channel estimation and data detection can be performed for the active users as those in conventional uplink cellular networks. However, as the user activity pattern cannot be known as prior, this scheme is unachievable in practice but can serve as a valuable performance upper bound.
\end{itemize}

\subsection{Results}
\begin{figure}[htpb]
\centering
\includegraphics[height=6cm,width=8cm]{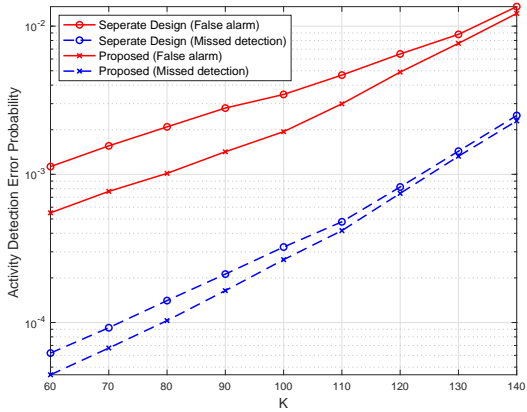}
\caption{Activity detection error probability vs. the number of active users.}
\label{4}
\end{figure}

We first evaluate the user activity detection error rate, including the false alarm and missed detection probabilities, depicted in Fig. \ref{4}. As seen from this figure, both the false alarm and missed detection probabilities increase with the number of active users. This is owing to the limited pilot resources available for user activity detection. Besides, for both the proposed framework and the separate design, it is observed that the false alarm probabilities dominate, which confirms the significance of the false alarm users to the overall activity detection accuracy. In addition, compared to the separate design, the proposed framework drastically reduces both types of activity detection error, which verifies its effectiveness in improving the activity detection performance by fully utilizing the sparsity pattern encoded in the received pilot and data signals. Moreover, we see that the performance improvement achieved by the proposed framework compared to the separate design is much more remarkable when $K$ is below 100, indicating that it is most effective when the traffic load in the systems ranges from light to medium.

\begin{figure}[htpb]
\centering
\includegraphics[height=6cm,width=8cm]{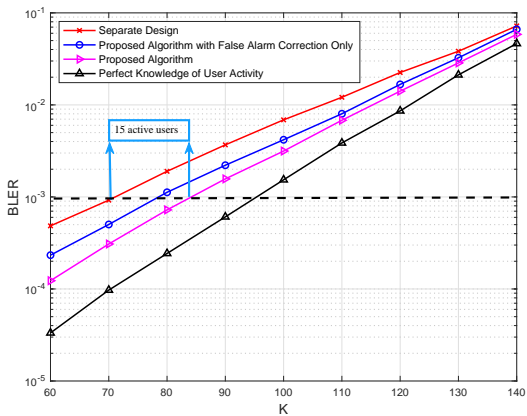}
\caption{BLER vs. the number of active users.}
\label{5}
\end{figure}

Next, we investigate the data error performance achieved by different algorithms and show the relationship between the block error rates (BLERs) and the number of active users in Fig. \ref{5}. Similar to the user activity detection error, the BLERs increase with the number of active users in the system. It is also noticed that the false alarm users have a significant impact on the BLER performance. For instance, when $K=100$, the proposed framework is able to reduce the BLER from $7\times10^{-3}$ to $4\times10^{-3}$ by invoking the false alarm correction only, while further applying the missed detection corrector only secures an extra 14.3\% BLER reduction. This matches the results in Fig. \ref{4}, where false alarm dominates the activity detection error. In addition, our proposed framework is able to support a substantially larger amount of active users compared to the baselines. For instance, if the BLER requirement is set to be $10^{-3}$, the proposed data-assisted user activity detection algorithm is capable of supporting fifteen additional users, which is a more than 20\%-improvement compared to the separate design. This again validates the superiority of the data-assisted design by fully exploiting the signal sparsity.

\section{Conclusions}
In this paper, we proposed a data-assisted user activity detection framework for massive random access. This framework effectively exploits the common sparsity pattern in both the received pilot and data signal, and thus boosts the performance of massive access for mMTC applications. Simulation results demonstrated that with the proposed framework, more than 20\% of active users can access the network with sufficient reliability. Based on this promising result, we advocate for a holistic approach on designing massive random access systems, by integrating the tasks of activity detection, channel estimation, and data detection and fully exploiting the available prior structure information. This calls for further investigations on efficient algorithms and theoretical analysis.

\end{document}